\documentclass{article}
\usepackage[final]{graphics}
\usepackage{amsfonts,amsbsy}

\font\tenmsa=msam10
\font\sevenmsa=msam7
\font\fivemsa=msam5
\font\tenmsb=msbm10
\font\sevenmsb=msbm7
\font\fivemsb=msbm5
\newfam\msafam
\newfam\msbfam
\textfont\msafam=\tenmsa  \scriptfont\msafam=\sevenmsa
\scriptscriptfont\msafam=\fivemsa
\textfont\msbfam=\tenmsb  \scriptfont\msbfam=\sevenmsb
\scriptscriptfont\msbfam=\fivemsb

\def\empile#1\over#2{\mathrel{\mathop{\kern 0pt#1}\limits_{#2}}}

\catcode`\@=11

\newcount\@tempcntc
\def\@citex[#1]#2{\if@filesw\immediate\write\@auxout{\string\citation{#2}}\fi
  \@tempcnta\z@\@tempcntb\m@ne\def\@citea{}\@cite{%
        \@for\@citeb:=#2\do%
    {\@ifundefined{b@\@citeb}%
        {\@citeo\@tempcntb\m@ne\@citea%
                \def\@citea{,\penalty\@m\ }{\bf ?}\@warning%
                {Citation `\@citeb' on page \thepage \space undefined}}%
        {\setbox\z@\hbox{\global\@tempcntc0\csname b@\@citeb\endcsname\relax}
     \ifnum\@tempcntc=\z@ \@citeo\@tempcntb\m@ne%
       \@citea\def\@citea{,\penalty\@m}%
       \hbox{\csname b@\@citeb\endcsname}%
     \else%
      \advance\@tempcntb\@ne%
      \ifnum\@tempcntb=\@tempcntc%
      \else\advance\@tempcntb\m@ne\@citeo%
      \@tempcnta\@tempcntc\@tempcntb\@tempcntc\fi\fi}}\@citeo}{#1}}%

\def\@citeo{\ifnum\@tempcnta>\@tempcntb\else\@citea
  \def\@citea{,\penalty\@m}%
  \ifnum\@tempcnta=\@tempcntb\the\@tempcnta\else
   {\advance\@tempcnta\@ne\ifnum\@tempcnta=\@tempcntb \else
\def\@citea{--}\fi
    \advance\@tempcnta\m@ne\the\@tempcnta\@citea\the\@tempcntb}\fi\fi}

\catcode`\@=12

\global\mathchardef\lesssim "142E

\newcommand{\slL}{\raise.15ex\hbox{$/$}\kern-.53em\hbox{$L$}}
\newcommand{\slP}{\raise.15ex\hbox{$/$}\kern-.53em\hbox{$P$}}
\newcommand{\slp}{\raise.1ex\hbox{$/$}\kern-.63em\hbox{$p$}}
\newcommand{\slq}{\raise.1ex\hbox{$/$}\kern-.63em\hbox{$q$}}
\newcommand{\slv}{\raise.1ex\hbox{$/$}\kern-.63em\hbox{$v$}}
\newcommand{\slR}{\raise.15ex\hbox{$/$}\kern-.53em\hbox{$R$}}
\newcommand{\slQ}{\raise.15ex\hbox{$/$}\kern-.53em\hbox{$Q$}}
\newcommand{\slK}{\raise.15ex\hbox{$/$}\kern-.53em\hbox{$K$}}
\newcommand{\slk}{\raise.15ex\hbox{$/$}\kern-.53em\hbox{$k$}}
\newcommand{\slSigma}{\raise.15ex\hbox{$/$}\kern-.53em\hbox{$\Sigma$}}
\newcommand{\slcalP}{\raise.15ex\hbox{$/$}\kern-.63em\hbox{$\cal P$}}
\newcommand{\slA}{\raise.15ex\hbox{$/$}\kern-.73em\hbox{$A$}}
\newcommand{\slbfA}{\raise.15ex\hbox{$/$}\kern-.73em\hbox{${\imb A}$}}
\newcommand{\slpartial}{\raise.15ex\hbox{$/$}\kern-.53em\hbox{$\partial$}}

\newcommand{\be}{\begin{equation}}
\newcommand{\ee}{\end{equation}}
\newcommand{\bea}{\begin{eqnarray}}
\newcommand{\ena}{\end{eqnarray}}

\def\build#1\over#2{\mathrel{\mathop{\kern 0pt#1}\limits_{#2}}}

\newcount\toto
\def\strip[hep-ph/0204#1]{#1}
\def\addoneto#1{\toto=#1\relax%
        \global\advance\toto by 1\relax%
        \the\toto}
\def\subtractoneto#1{\toto=#1\relax%
        \global\advance\toto by -1\relax%
        \the\toto}

\font\tenimbf=cmmib10 at 10pt
\font\sevenimbf=cmmib10 at 7pt
\font\fiveimbf=cmmib10 at 5pt
\newfam\imbf
\textfont\imbf=\tenimbf
\scriptfont\imbf=\sevenimbf
\scriptscriptfont\imbf=\fiveimbf
\def\imb{\fam\imbf\tenimbf}

\begin{document}
\title{\bf{A simple sum rule for the thermal gluon spectral function and applications}}
\author{P.~Aurenche$^{(1)}$, F.~Gelis$^{(2)}$, H.~Zaraket$^{(3)}$}
\maketitle
\begin{center}
\begin{enumerate}
\item Laboratoire d'Annecy-le-Vieux de Physique Th\'eorique,\\
Chemin de Bellevue, B.P. 110,\\
74941 Annecy-le-Vieux Cedex, France
\item Laboratoire de Physique Th\'eorique,\\
B\^at. 210, Universit\'e Paris XI,\\
91405 Orsay Cedex, France
\item Physics Department and Winnipeg Institute
for Theoretical Physics,\\
University of Winnipeg,\\
Winnipeg, Manitoba R3B 2E9, Canada
\end{enumerate}
\end{center}

\begin{abstract}

  In this paper, we derive a simple sum rule satisfied by the gluon
  spectral function at finite temperature. This sum rule is useful in
  order to calculate exactly some integrals that appear frequently in
  the photon or dilepton production rate by a quark gluon plasma.
  Using this sum rule, we rederive simply some known results and
  obtain some new results that would be extremely difficult to justify
  otherwise. In particular, we derive an exact expression for the
  collision integral that appears in the calculation of the
  Landau-Pomeranchuk-Migdal effect.
\end{abstract}
\vskip 4mm
\centerline{\hfill LAPTH-909/02, LPT-ORSAY-02/27}

\section{Introduction}
Photon production is considered to be a very interesting signal of the
formation of a quark gluon plasma in heavy ion collisions
\cite{PeitzT1,Aggara1,Aggara2,Sriva1,SrivaS1,SollfHKRP1,HuoviRR1}. Indeed,
because of their very weak coupling to matter, photons (and more
generally any electromagnetic probe) have a large mean free path which
enables them to escape without reinteractions from a medium the size of
which is at best a few tens of fermis.

On the theoretical side, the calculation of the photon and dilepton rate
is performed under the hypothesis of local equilibrium, i.e. one
calculates a rate (number of photons produced per unit time and per unit
volume) using thermal field theory in equilibrium, and then plugs this
rate into some hydrodynamical model
\cite{PeitzT1,Sriva1,SrivaS1,SollfHKRP1,HuoviRR1} which describes the
system by dividing it in cells where a local equilibrium is assumed and
by assigning a local temperature and fluid 4-velocity to each cell. Such
a description is consistent as long as the photon formation time is small
compared to the typical size of the cells in which an approximate
equilibrium is realized \cite{Gelis11,GelisSS1}.

\begin{figure}
\centerline{\resizebox*{!}{3cm}{\includegraphics{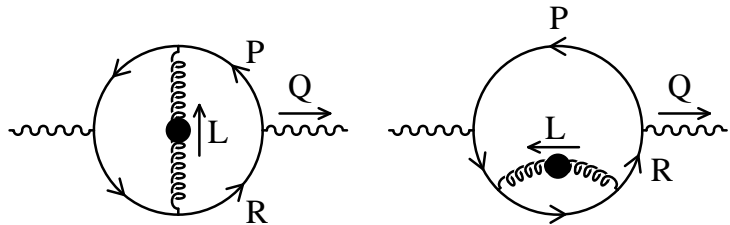}}}
\caption{\label{fig:diagrams} The two-loop diagrams contributing 
  to photon and dilepton production.}
\end{figure}

\begin{figure}
\centerline{\resizebox*{!}{3.5cm}{\includegraphics{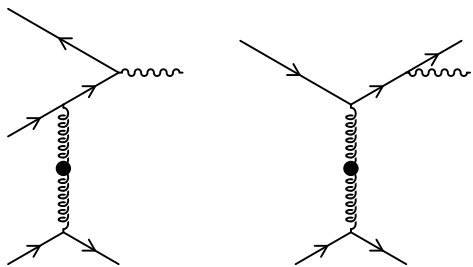}}}
\caption{\label{fig:processes} Important processes contained in the
above 2-loop diagrams.}
\end{figure}

Thermal field theory calculations of photon and dilepton production
rates have been performed a long time ago
\cite{AltheA1,AltheR1,AltheB2,GabelGP1}, and have been reassessed
under the new light shaded by the concept of hard thermal loops (HTL -
\cite{Pisar6,BraatP1,BraatP2,FrenkT1,FrenkT2})
\cite{BraatPY1,BaierNNR1,KapusLS1,BaierPS1,AurenBP1,Niega6}. In this
context, it has been found that some $2\to 3$ and $3\to 2$ processes
which appear only in 2-loop diagrams
\cite{AurenGKP1,AurenGKP2,AurenGKZ1,AurenGKZ2} (like bremsstrahlung,
as well as the process $q\bar{q}\{q,g\}\to \gamma\{q,g\}$ which can be
deduced from bremsstrahlung by crossing symmetry - see figures
\ref{fig:diagrams} and \ref{fig:processes}) are also important.
These processes are in fact enhanced by a strong sensitivity to the
forward emission of the photon, due to a collinear singularity
regularized by a thermal mass of order $gT$ where $g$ is the gauge
coupling. In particular, the process on the left of figure
\ref{fig:processes} has been found to enhance considerably the rate of
hard photons \cite{AurenGKZ1}.  This collinear enhancement mechanism has
also been shown to play a role in multi-loop diagrams belonging to the
class of ladder corrections and self-energy-corrections
\cite{LebedS1,LebedS2,AurenGZ2,Gelis11,ArnolMY1,ArnolMY2}, and a
resummation of this family of diagrams has been carried out in
\cite{ArnolMY1,ArnolMY2}. The effect of this resummation, also known as
Landau-Pomeranchuk-Migdal (LPM) effect \cite{LandaP1,LandaP2,Migda1},
leads to a small reduction (by about 25\% for photons in the range
interesting for phenomenology) of the photon rate.

However, the strength of the LPM suppression decreases as one
increases the invariant mass of the photon since the photon mass helps
to regularize the collinear singularities. It is therefore expected
that the dilepton rate can be accounted for by limiting oneself to a
2-loop calculation, and that the production of hard dileptons of
intermediate invariant mass is dominated by the process shown on the
left of figure \ref{fig:processes}. Under this assumption, the
calculation of \cite{AurenGKZ1} has been extended recently to the case
where the photon mass cannot be neglected any longer, with emphasis on
the ``off-shell annihilation'' process that was already dominant for
hard photons \cite{AurenGZ3}. In this paper, we found a simple
generalization to the case of massive photons of the formula known for
the imaginary part of the real photon polarization tensor. This
formula reads:
\begin{eqnarray}
&&\!\!\!\!{\rm Im}\,\Pi_{_{R}}{}_\mu^\mu(Q)\approx 
-{{e^2 g^2 N_c C_{_{F}}}\over{2\pi^4}}{T\over{q_0^2}}
\int\limits_{-\infty}^{+\infty}dp_0[n_{_{F}}(r_0)-n_{_{F}}(p_0)]\nonumber\\
&&\quad\times\left\{(p_0^2+r_0^2)(J_{_{T}}-J_{_{L}})+2{{Q^2 p_0r_0+M_\infty^2(p_0^2+r_0^2)}\over{M_{\rm
eff}^2}} (K_{_{T}}-K_{_{L}})\right\}\, .
\label{eq:rate-final}
\end{eqnarray}
with $r_0\equiv p_0+q_0$ and\footnote{Let us recall here that $M_{\rm
    eff}^2$ can become negative if $Q^2>4M_\infty^2$. Those
  definitions are only appropriate for $M_{\rm eff}^2>0$, because they
  have been obtained by rescaling the transverse momentum $l_\perp$ of
  the gluon by writing $l_\perp^2\equiv M_{\rm eff}^2 w$. However, it
  was argued in \cite{AurenGZ3} that the correct result for $M_{\rm
    eff}^2<0$ can be obtained as the real part of the analytic
  continuation of the result for $M_{\rm eff}^2>0$. Most of this
  paper deals with the case of a positive $M_{\rm eff}^2$.}
\begin{eqnarray}
&&\!\!\!\!\!J_{_{T,L}}\equiv M_{\rm eff}^2 \int\limits_{0}^{1}\! {{dx}\over{x}} 
{\rm
Im}\,\Pi_{_{T,L}}(x) \!\int\limits_{0}^{+\infty}\!\! dw
{{\sqrt{w/(w+4)} {\rm tanh}^{-1}\sqrt{w/(w+4)}}\over
{(M_{\rm eff}^2w+{\rm Re}\,\Pi_{_{T,L}}(x))^2+({\rm Im}\,\Pi_{_{T,L}}(x))^2}}\;
,\nonumber\\
&&\!\!\!\!\!K_{_{T,L}}\equiv M_{\rm eff}^2 \int\limits_{0}^{1}\! {{dx}\over{x}} 
{\rm
Im}\,\Pi_{_{T,L}}(x) \!\int\limits_{0}^{+\infty}\!\! {{dw}\over w}
{{\sqrt{w/(w+4)} {\rm tanh}^{-1}\sqrt{w/(w+4)}-w/4}\over
{(M_{\rm eff}^2w+{\rm Re}\,\Pi_{_{T,L}}(x))^2+({\rm Im}\,\Pi_{_{T,L}}(x))^2}}\;
,\nonumber\\
&&
\end{eqnarray}
where the $\Pi_{_{T,L}}$ are the transverse and longitudinal self-energies
resummed on the gluon propagator, and where we denote
\begin{equation}
M_{\rm eff}^2\equiv M_\infty^2+{{Q^2}\over{q_0^2}}p_0r_0\; ,
\end{equation}
with $M_\infty$ the thermal mass of a hard quark
($M_\infty^2=g^2C_{_{F}}T^2/4$). The functions $\Pi_{_{T,L}}$ are
the usual transverse and longitudinal HTL gluon self-energies:
\begin{eqnarray}
&\Pi_{_{T}}(L)=3 m_{\rm g}^2&\left[ {{x^2}\over
2}+{{x(1-x^2)}\over{4}}\ln\left({{x+1}\over{x-1}}\right)
\right]\nonumber\\ 
&\Pi_{_{L}}(L)=3 m_{\rm g}^2&\left[
(1-x^2)-{{x(1-x^2)}\over{2}}\ln\left({{x+1}\over{x-1}}\right)
\right]\; ,\nonumber\\ &&
\label{eq:gluonself}
\end{eqnarray}
where we denote $x\equiv l_0/l$ and where $m_{\rm g}^2\equiv g^2 T^2 [N_c+
N_{_{F}}/2]/9$ is the gluon thermal mass in a $SU(N_c)$ gauge theory
with $N_{_{F}}$ flavors.

Therefore, in addition to the function $J_{_{T,L}}$ already introduced
in the case of quasi-real photons\footnote{The term in $M_\infty^2
  K_{_{T,L}}$ was forgotten in \cite{AurenGKP2}. It comes from the HTL
  correction to the $\gamma q\bar{q}$ vertex. This vertex correction
  was also neglected in \cite{ArnolMY1,ArnolMY2}, without any damage
  to this approach since it affects only the component $\Pi_{zz}$ of
  the polarization tensor, while only the transverse components are
  calculated in these papers (see \cite{AurenGZ3} for more details on
  this issue).}, we need two new functions $K_{_{T,L}}$ for the term
proportional to the photon invariant mass squared $Q^2$. All are
dimensionless functions of the ratio of $M_{\rm eff}$ to the plasmon
mass $m_{\rm g}$ which appears as a prefactor in the self-energies
$\Pi_{_{T,L}}$. Up to now, the $J_{_{T,L}}$ and $K_{_{T,L}}$ have only
been evaluated numerically (\cite{AurenGKP2,AurenGKZ1}, with a mistake
corrected by \cite{Mohan1,Mohan2,SteffT1}), which is sufficient for
the case of real photons since in this case they are fixed numbers
that depend only on the number of colors and flavors but not on
kinematical parameters ($M_{\rm eff}^2=M_\infty^2$).  However, the
cost of this procedure increases significantly in the case of virtual
photons since the value of $M_{\rm eff}^2$ depends on the invariant
mass $Q$, energy $q_0$ and quark energy $p_0$. In addition, obtaining
asymptotic limits is far from trivial at this point\footnote{Some very
  partial asymptotic results have been obtained in \cite{AurenGKP2}
  for $J_{_{T,L}}$.}.

The aim of the present paper is to derive some analytical results
regarding those functions.  We first show how the integral over the
variable $x$ can be performed exactly in the functions $J_{_{T,L}}$
and $K_{_{T,L}}$ by means of a simple sum-rule (section
\ref{sec:sum-rule}). This leads to either a very simple integral
representation of these functions or even to closed formulas in terms
of dilogarithms (section \ref{sec:JK}). Thanks to these results, we
can easily study the asymptotic properties of the functions
$J_{_{T,L}}$ and $K_{_{T,L}}$. Non trivial asymptotic expansions are
obtained, which would have been very difficult to obtain otherwise
(section \ref{sec:JK_asympt}). In section \ref{sec:beyond-HTL}, we
show that the above analytic results can also give some insight on the
fact that the processes of figure \ref{fig:processes} depend only on
parameters like the gluon screening masses and the hard quark thermal
mass, in a generic model where the quark gluon plasma is described as
a gas of quasi-particles.
Finally, we show that one can also calculate analytically the
collision integral that appears in the resummation of ladder diagrams
\cite{ArnolMY1,ArnolMY2} (section \ref{sec:resum}).  In appendix
\ref{app:F-asympt}, we derive the asymptotic behavior of a function
introduced at an intermediate stage.  In appendix \ref{app:B}, we prove
analytically an anecdotical property which was first noticed
numerically \cite{Kobes4}: the integrals $J_{_{T}}$ and $J_{_{L}}$ are
exactly opposite if $M_\infty=m_{\rm g}$ and $Q^2=0$.

\section{Derivation of the sum-rule}
\label{sec:sum-rule}
We want to calculate the integral:
\begin{equation}
f(z)\equiv\int_{0}^{1}{{dx}\over{x}} 
{{2{\rm Im}\,\Pi(x)}\over{(z+{\rm Re}\,\Pi(x))^2+({\rm
Im}\,\Pi(x))^2}}\; ,
\end{equation}
for a positive $z$, where $\Pi(x)$ is some self-energy depending only
on $x\equiv k_0/k$ as is the case for instance with the HTL gluonic
self-energy. The factor $1/x$ comes from a Bose-Einstein factor in the
soft approximation $dk_0 n_{_{B}}(k_0)\approx T dk_0 /k_0 = T dx/x$.
The first step in this calculation is to rewrite it as
\begin{equation}
f(z)=\int_{0}^{1}{{dx}\over{x}} (1-x^2)
{{2{\rm Im}\,\overline{\Pi}(x)}\over{(z(x^2-1)-{\rm Re}\,\overline{\Pi}(x))^2+({\rm
Im}\,\overline{\Pi}(x))^2}}\; ,
\end{equation}
where we define $\overline{\Pi}(x)\equiv (1-x^2)\Pi(x)$. Interpreting
now $z$ ($z>0$) as the square of a three-momentum $k$ and $x$ as the
ratio $x=k_0/k$, we have
\begin{eqnarray}
{{-2{\rm Im}\,\overline{\Pi}(x)}\over{(z(x^2-1)-{\rm Re}\,\overline{\Pi}(x))^2+({\rm
Im}\,\overline{\Pi}(x))^2}}&&=2{\rm Re}\,
{i\over{k_0^2-k^2-\overline{\Pi}(k_0,k)+ik_0\epsilon}}\nonumber\\
&&\equiv\overline{\rho}(k_0,k)\; .
\label{eq:def-rho}
\end{eqnarray}
Note that this function is nothing but the spectral function
$\overline{\rho}(k_0,k)$ associated with the ``propagator'' appearing
in the right hand side.  This is what enables us to relate the
integral
\begin{equation}
f(z)=-\int\limits_0^{1}{{dx}\over x} (1-x^2) \overline{\rho}(\sqrt{z}x,\sqrt{z})
\end{equation}
to the spectral representation of this propagator. Indeed, it is known
that the resummed propagator $i/(K^2-\overline{\Pi}(K))$ is related to
its spectral function $\overline{\rho}(k_0,k)$ via the following
spectral representation \cite{LandsW1,Gelis7}:
\begin{equation}
{i\over{k_0^2-k^2-\overline{\Pi}(k_0/k)+ik_0\epsilon}}
=\int_{0}^{+\infty} {{dE}\over{\pi}} E \overline{\rho}(E,k)
{{i}\over{k_0^2-E^2+ik_0\epsilon}}\; .
\end{equation}
Taking the real part of this identity\footnote{Note that the spectral
function $\overline{\rho}(k_0,k)$ is by definition a real function.}, one recovers
the definition Eq.~(\ref{eq:def-rho}) of the spectral function. Taking
its imaginary part and denoting $E\equiv k x$ and $k_0=k y$, one
obtains the following non-trivial integral:
\begin{equation}\
\int_{0}^{+\infty} {{dx}\over{\pi}} x\overline{\rho}(kx,k) {1\over{y^2-x^2}}
={{k^2(y^2-1)-{\rm Re}\,\overline{\Pi}(y)}
\over{(k^2(y^2-1)-{\rm Re}\,\overline{\Pi}(y))^2+({\rm Im}\,\overline{\Pi}(y))^2}}\; .
\end{equation} 
Taking the limit $y\to\infty$, we find
\begin{equation}
\int_{0}^{+\infty} {{dx}\over{\pi}} x\overline{\rho}(kx,k)=
{1\over{k^2-\lim_{y\to\infty}{\rm
Re}\,\overline{\Pi}(y)/y^2}}={1\over{k^2+{\rm Re}\,\Pi(\infty)}}\; .
\end{equation}
Having in mind that $\Pi(y)$ is a gluonic self-energy obtained from the
hard thermal loop approximation, its imaginary part vanishes if $y\ge 1$
and therefore does not contribute in the limit $y\to
\infty$. Alternatively, taking the limit $y\to 0$ and assuming similarly
that ${\rm Im}\,\Pi(y=0)=0$, we obtain:
\begin{equation}
\int_{0}^{+\infty} {{dx}\over{\pi}} {1\over x}\overline{\rho}(kx,k)=
{1\over{k^2+{\rm Re}\,\Pi(0)}}\; .
\end{equation}
We can combine these two relations into
\begin{equation}
\int_{0}^{+\infty} {{dx}\over{x}} (1-x^2) \overline{\rho}(kx,k)=\pi\left[
{1\over{k^2+{\rm Re}\,\Pi(0)}}
-
{1\over{k^2+{\rm Re}\,\Pi(\infty)}}
 \right]\; .
\label{eq:total}
\end{equation}

In order to obtain from there the function $f(z)$, we need to subtract
the contribution coming from $x$ between $1$ and $+\infty$.
Fortunately, since ${\rm Im}\,\Pi(x)=0$ for $x\ge 1$, the contribution
in this range comes only from the poles of the propagator
$i/(K^2-\overline{\Pi}(K))$, via the formula\footnote{If ${\rm Im}\,\overline{\Pi}(x)=0$, then $\overline{\rho}(kx,k)=2\pi\delta(k^2(x^2-1)-\overline{\Pi}(x))$.}
\begin{equation}
\int_{1}^{+\infty} {{dx}\over x}(1-x^2)\overline{\rho}(kx,k)=\pi\sum_{{\rm poles\ } x_i}
{{Z(x_i)}\over{k^2}} {{1-x_i^2}\over{x_i^2}}\; ,
\label{eq:poles}
\end{equation}
where $Z(x_i)$ is the residue of the propagator at the corresponding
pole.
For the above propagator, the equation that determines the poles is
\begin{equation}
k^2(x^2-1)={\rm Re}\,\overline{\Pi}(x)=(1-x^2){\rm Re}\,\Pi(x)\; .
\end{equation}

This ``dispersion equation'' has a trivial solution $x=1$, which does
not contribute when plugged in Eq.~(\ref{eq:poles}) because the other
factors in the integrand vanish if $x=1$. Any non trivial pole would be
a solution of the equation
\begin{equation}
k^2+{\rm Re}\,\Pi(x)=0\;,
\end{equation}
but under the reasonable assumption that the resummation of the
self-energy $\Pi(x)$ leads to well behaved quasi-particles (i.e. that
the equation $k_0^2-k^2=\Pi(k_0/k)$ has a solution for every value of
$k_0/k$ larger than $1$), we have ${\rm Re}\,\Pi(x) \ge 0$ for $x>1$ and
therefore there are no additional poles. As a consequence, the integral
of Eq.~(\ref{eq:total}) does not receive any contribution from the range
$x\in[1,+\infty]$, and we can write directly a closed expression for the
function $f(z)$:
\begin{equation}
f(z)=\!\int_{0}^{1}\!{{dx}\over{x}} 
{{2{\rm Im}\,\Pi(x)}\over{(z+{\rm Re}\,\Pi(x))^2+({\rm
Im}\,\Pi(x))^2}}=\pi\left[{1\over{z+{\rm Re}\,\Pi(\infty)}}
\!-\!
{1\over{z+{\rm Re}\,\Pi(0)}} \right]\, .
\label{eq:fz}
\end{equation}
This is the basic sum-rule from which we are going to derive some
results regarding photon production by a quark-gluon plasma. The
validity of this result can also be checked numerically.

It may be useful to recall that, even if the derivation has been made
having in mind a hard thermal loop for the self-energy $\Pi$, this
result has a broader range of validity. In fact, it is valid for any
self-energy satisfying the following assumptions:
\begin{enumerate}
\item $\Pi$ depends only on $x\equiv k_0/k$
\item ${\rm Im}\,\Pi(x=0)=0$
\item ${\rm Im}\,\Pi(x)=0$ if $x\ge 1$
\item ${\rm Re}\,\Pi(x)\ge 0$ if $x\ge 1$.
\end{enumerate}
Note that the condition (1) can in fact be relaxed since what is
done here can be reproduced if the self-energy depends separately on
$k$ and $x$, the only difference being that the result would depend on
$k$. The condition (2) is always true. Conditions (3) and (4) depend on
the nature of the resummation under consideration, but are reasonable
approximations in any system of well defined quasiparticles.

In \cite{AurenGKP2}, a formula for the photon rate based on sum rules
was also presented. However, in this paper, the use of sum rules led
only to a very complicated result (involving explicitly the gluon
dispersion equations as well as the residues of the gluon poles) that
was not useful for any practical purpose. By comparing the two methods,
we can trace the simplification achieved in the present paper to a
different choice for the integration variables. Indeed, in
\cite{AurenGKP2} the sum rules were applied to an integral over the
variables $x\equiv l_0/l$, $l=|{\imb l}|$ (where $L$ is the momentum of
the exchanged gluon), while in the present approach, we take as
independent integration variables $x$ and $w\equiv -L^2/M_{\rm eff}^2=l^2(1-x^2)/M_{\rm eff}^2$. It
appears that trading $l$ in favor of $w$ before using sum rules to
perform the integral over $x$ leads to a dramatic simplification of the
result, because some non trivial parts of the $x$ dependence get
absorbed in the new variable $w$. There are in fact many sum rules
satisfied by the HTL spectral functions. The interested reader may
find other examples in \cite{BellaR2,Reyna1,VandeO1}.

\section{Expression of $J_{_{T,L}}$ and $K_{_{T,L}}$}
\label{sec:JK}
Thanks to the formula derived in the previous section, one can first
simplify the functions $J_{_{T,L}}$ and $K_{_{T,L}}$ so that we have
only one-dimensional integrals over the variable $w$:
\begin{eqnarray}
&&J_{_{T,L}}={\pi\over 2}
\int\limits_{0}^{+\infty} dw \sqrt{w\over{w+4}}
{\rm tanh}^{-1} \sqrt{w\over{w+4}} \nonumber\\
&&\qquad\qquad\qquad\times\left[
{1\over{w+{{{\rm Re}\,\Pi_{_{T,L}}(\infty)}\over{M_{\rm eff}^2}}}}
-
{1\over{w+{{{\rm Re}\,\Pi_{_{T,L}}(0)}\over{M_{\rm eff}^2}}}}
\right]\; ,\nonumber\\
&&K_{_{T,L}}={\pi\over 2}
\int\limits_{0}^{+\infty} {{dw}\over{w}} \left[\sqrt{w\over{w+4}}
{\rm tanh}^{-1} \sqrt{w\over{w+4}} -{w\over 4}\right]\nonumber\\
&&\qquad\qquad\qquad\times\left[
{1\over{w+{{{\rm Re}\,\Pi_{_{T,L}}(\infty)}\over{M_{\rm eff}^2}}}}
-
{1\over{w+{{{\rm Re}\,\Pi_{_{T,L}}(0)}\over{M_{\rm eff}^2}}}}
\right]\; .
\end{eqnarray}

At this point, using the change of variable $u\equiv\sqrt{w/(w+4)}$, we
can write in a simpler way the following elementary integrals:
\begin{eqnarray}
&&\int\limits_{0}^{+\infty} dw \sqrt{w\over{w+4}}
{\rm tanh}^{-1} \sqrt{w\over{w+4}} \left[{1\over{w}}-{1\over{w+a}}\right]=
2F\Big({4\over a}\Big)\; ,\nonumber\\
&&\int\limits_{0}^{+\infty} {{dw}\over{w}} \left[\sqrt{w\over{w+4}}
{\rm tanh}^{-1} \sqrt{w\over{w+4}} -{w\over 4}\right]
\left[{1\over{w}}-{1\over{w+a}}\right]\nonumber\\
&&\qquad\qquad\qquad\qquad={1\over
4}\ln\Big({1\over a}\Big)+{1\over 2}
-{2\over a}F\Big({4\over a}\Big)\; ,
\end{eqnarray}
where we define the function
\begin{equation}
F(x)\equiv \int\limits_{0}^{1} du {{{\rm
tanh}^{-1}(u)}\over{(x-1)u^2+1}}\; .
\label{eq:F-def}
\end{equation}
Recalling now the following properties of the HTL self-energy of the
gluon \cite{Pisar6}
\begin{eqnarray} 
&&{\rm Re}\,\Pi_{_{T,L}}(\infty)=m_{\rm g}^2\nonumber\\
&&{\rm Re}\,\Pi_{_{T}}(0)=0\nonumber\\
&&{\rm Re}\,\Pi_{_{L}}(0)=3 m_{\rm g}^2
\end{eqnarray}
where $m_{\rm g}$ is the plasmon mass, we easily obtain the following
expressions in terms of the function $F(x)$:
\begin{eqnarray}
&&J_{_{T}}=-\pi F\Big({{4M_{\rm eff}^2}\over{m_{\rm g}^2}}\Big)\; ,\nonumber\\
&&J_{_{L}}=\pi\left[ F\Big({{4M_{\rm eff}^2}\over{3m_{\rm g}^2}}\Big) - 
F\Big({{4M_{\rm eff}^2}\over{m_{\rm g}^2}}\Big)
\right]\; ,\nonumber\\
&&K_{_{T}}=\pi\left[{{M_{\rm eff}^2}\over{m_{\rm g}^2}} F\Big({{4M_{\rm
eff}^2}\over{m_{\rm g}^2}}\Big)
-{1\over 4} - {1\over 8}\ln\Big(
{{M_{\rm eff}^2}\over{m_{\rm g}^2}}
\Big)\right]\; ,\nonumber\\
&&K_{_{L}}=\pi\left[
-{1\over 8}\ln(3)+{{M_{\rm eff}^2}\over{m_{\rm g}^2}}\left(
F\Big({{4M_{\rm
eff}^2}\over{m_{\rm g}^2}}\Big)
-{1\over 3}F\Big({{4M_{\rm
eff}^2}\over{3m_{\rm g}^2}}\Big)
\right)
\right]\; .
\label{eq:KL-final}
\end{eqnarray}
Therefore, those results demonstrate that in order to study the
properties of the functions $J_{_{T,L}}$ and $K_{_{T,L}}$, one needs
only to study the properties of the much simpler function $F(x)$. In
fact, it is even possible to write the function $F(x)$ in closed form in
terms of dilogarithms\footnote{Explicitly, we have:
\begin{eqnarray}
&&F(x)={1\over {4ip}}\left[{\rm Li}_2\left({{2p}\over{p+i}}\right)
-2{\rm Li}_2\left({{p}\over{p+i}}\right)
-{\rm Li}_2\left({{2p}\over{p-i}}\right)
+2{\rm Li}_2\left({{p}\over{p-i}}\right)\right.\nonumber\\
&&\qquad\left.
+2\ln(2)\ln\left({{i-p}\over{i+p}}\right)\right]\; ,\qquad{\rm
with}\quad p\equiv\sqrt{x-1}\quad{\rm and}\quad {\rm Li}_2(x)\equiv\sum_{n=1}^{+\infty}{{x^n}\over{n^2}}\; .
\end{eqnarray}}. We are not going to make use of this possibility
here since it is simpler to keep $F(x)$ in its integral form given by
Eq.~(\ref{eq:F-def}).

One must stress the fact that all these functions depend only on the
ratio of two masses. In the case of real photons ($Q^2=0$), we have in
addition $M_{\rm eff}^2=M_\infty^2$ and for $N_c=3$ colors, we can write:
\begin{equation}
{{4M_\infty^2}\over{3m_{\rm g}^2}}={8\over{6+N_{_{F}}}}\; ,
\end{equation}
i.e. the temperature and strong coupling constant also drop out of this ratio.
It appears that there is an additional (and purely accidental)
simplification for $N_{_{F}}=2$ flavors: in this case, the differences
$J_{_{L}}-J_{_{T}}$ and $K_{_{L}}-K_{_{T}}$ can be expressed in a very
simple fashion:
\begin{eqnarray}
&&J_{_{L}}-J_{_{T}}\Big|_{N_c=3,N_{_{F}}=2}=\pi\ln(2)\; ,\nonumber\\
&&K_{_{L}}-K_{_{T}}\Big|_{N_c=3,N_{_{F}}=2}={\pi\over 4}(1-2\ln(2))\; .
\end{eqnarray}
 For the case of
$N_{_{F}}=3$ flavors, the results are less simple, but one can still
obtain explicit expressions:
\begin{eqnarray}
&&J_{_{L}}-J_{_{T}}\Big|_{N_c=3,N_{_{F}}=3}=\pi\Big[{{\pi^2}\over{8}}
-{{33}\over 8}\ln^2(2)+3\ln(2)\ln(3)\nonumber\\
&&\qquad\qquad\qquad\qquad\qquad\quad
-{3\over 2}{\rm Li}_2\Big({3\over 4}\Big)
-{3\over 2}{\rm Li}_2\Big(-{1\over 2}\Big)\Big]\; ,\nonumber\\
&&K_{_{L}}-K_{_{T}}\Big|_{N_c=3,N_{_{F}}=3}=\pi\Big[
{1\over 4}-{{\pi^2}\over{36}}+{{\ln(2)}\over 8}-{{\ln(3)}\over 4}
+{{11}\over{12}}\ln^2(2)\nonumber\\
&&\qquad\qquad\qquad\qquad\qquad\quad
-{2\over 3}\ln(2)\ln(3)
+{1\over 3}{\rm Li}_2\Big({3\over 4}\Big)
+{1\over 3}{\rm Li}_2\Big(-{1\over 2}\Big)
\Big]\; .\nonumber\\
&&
\end{eqnarray}
Had these exact formulas been known, the confusion due to the
erroneous factor $4$ in the numerical evaluation of these coefficients
in \cite{AurenGKZ1} would have been avoided \cite{SteffT1}.

\section{Asymptotic behavior of $J_{_{T,L}}$ and $K_{_{T,L}}$}
\label{sec:JK_asympt}
\subsection{Limit $M_{\rm eff}\ll m_{\rm g}$}
Using the above results, we can recover in a rather simple and elegant
way all the asymptotic limits given in \cite{AurenGKP2} for
$J_{_{T,L}}$, as well as the limits used for $K_{_{T,L}}$ in order to
obtain the behavior of 2-loop dilepton production near the threshold
$Q^2=4M_\infty^2$ (a region which is dominated by small values of
$M_{\rm eff}^2)$ \cite{AurenGZ3}.

To that effect, we need only the behavior of $F(x)$ when $x\to 0$. This
is derived in the appendix \ref{app:F-asympt}, where we prove that:
\begin{equation}
F(x)\empile{=}\over{x\to 0^+} {1\over 8} \ln^2\Big({4\over x}\Big)
+{{\pi^2}\over{24}}+{\cal O}\Big(x
\ln^2(1/x)\Big)\; .
\label{eq:F-small-x}
\end{equation}
Thanks to this formula, a trivial calculation gives\footnote{These
leading log formulas are not affected by the fact that the asymptotic
expansion of $F(x)$ is slightly modified if $x$ approaches $0$ with
negative values (see Eq.~(\ref{eq:F-small-neg-x})).}
\begin{eqnarray}
&&J_{_{T}}\empile{\approx}\over{M_{\rm eff}\ll m_{\rm g}} -{\pi\over 8}\ln^2
\Big({{m_{\rm g}^2}\over{M_{\rm eff}^2}}\Big)\; ,\nonumber\\
&&J_{_{L}}\empile{\approx}\over{M_{\rm eff}\ll m_{\rm g}} {{\pi\ln(3)}\over 4}
\ln\Big({{m_{\rm g}^2}\over{M_{\rm eff}^2}}\Big)\; ,\nonumber\\
&&K_{_{T}}\empile{\approx}\over{M_{\rm eff}\ll m_{\rm g}} {\pi\over 8}
\left[\ln\Big({{m_{\rm g}^2}\over{M_{\rm eff}^2}}\Big)-2\right]\; ,\nonumber\\
&&K_{_{L}}\empile{\approx}\over{M_{\rm eff}\ll m_{\rm g}}-{{\pi\ln(3)}\over 8}\; .
\end{eqnarray}
These relations in fact go well beyond the results for $J_{_{T,L}}$ obtained in
\cite{AurenGKP2}, since in this new approach we obtain for free the
prefactor of the leading term and we could even have calculated some
subleading terms, down to the constant term.

\subsection{Limit $M_{\rm eff}\gg m_{\rm g}$}
Using Eqs.~(\ref{eq:KL-final}), it is also very simple to obtain the
behavior of the functions $J_{_{T,L}}$ and $K_{_{T,L}}$ in the opposite
limit where $M_{\rm eff}\gg m_{\rm g}$. In order to do that, we need to know
the behavior of $F(x)$ for large values of $x$. The following formula is
also derived in appendix \ref{app:F-asympt}:
\begin{equation}
F(x)\empile{=}\over{x\gg 1}
{{\ln(x)}\over{2x}}+{{1-\ln(2)}\over{x}}+{{\ln(x)}\over{3x^2}}+{{5-6\ln(2)}\over{9x^2}}+{\cal
O}\Big({{\ln(x)}\over{x^3}}
\Big)\; .
\end{equation}
From this formula, it is easy to obtain
\begin{eqnarray}
&&J_{_{T}}\empile{\approx}\over{M_{\rm eff}\gg m_{\rm g}} -{\pi\over 8}
{{m_{\rm g}^2}\over{M_{\rm eff}^2}} 
\ln\Big({{M_{\rm eff}^2}\over{m_{\rm g}^2}}\Big)\; ,\nonumber\\
&&J_{_{L}}\empile{\approx}\over{M_{\rm eff}\gg m_{\rm g}} {\pi\over 4}
{{m_{\rm g}^2}\over{M_{\rm eff}^2}} 
\ln\Big({{M_{\rm eff}^2}\over{m_{\rm g}^2}}\Big)\approx -2J_{_{T}}\; ,\nonumber\\
&&K_{_{T}}\empile{\approx}\over{M_{\rm eff}\gg m_{\rm g}}
{\pi\over{48}}{{m_{\rm g}^2}\over{M_{\rm eff}^2}}
\ln\Big({{M_{\rm eff}^2}\over{m_{\rm g}^2}}\Big)\; ,\nonumber\\
&&K_{_{L}}\empile{\approx}\over{M_{\rm eff}\gg m_{\rm g}}
-{\pi\over{24}}{{m_{\rm g}^2}\over{M_{\rm eff}^2}}
\ln\Big({{M_{\rm eff}^2}\over{m_{\rm g}^2}}\Big)\approx -2K_{_{T}}\; .
\label{eq:Mggm}
\end{eqnarray}

\section{Beyond the HTL approximation}
\label{sec:beyond-HTL}
In section \ref{sec:sum-rule}, we mentioned the fact that the sum rule
in Eq.~(\ref{eq:fz}) is in fact valid for a gluon self-energy $\Pi$ more
general than the standard case of hard thermal loops. Assuming it can be
applied, we
see that the result depends only on the four numbers ${\rm
Re}\,\Pi_{_{T,L}}(\infty)$ and ${\rm Re}\,\Pi_{_{T,L}}(0)$.  The first
two are the plasmon mass (longitudinal) and the mass of the transverse
gluon at zero momentum, and can be shown to be equal thanks to
Slavnor-Taylor identities \cite{KobesKR1,BraatP3,DirksNO1}. Physically,
this property means that there is no way to distinguish transverse and
longitudinal modes for a particle at rest. Therefore, we need only to
introduce one plasmon mass:
\begin{equation}
{\rm Re}\,\Pi_{_{T}}(\infty)={\rm Re}\,\Pi_{_{L}}(\infty)\equiv m_P^2\; .
\end{equation}
The quantities on the second line are squares of the screening masses
for the transverse and longitudinal static gluon exchanges. The
longitudinal screening mass is the familiar Debye mass:
\begin{equation}
m_D^2\equiv {\rm
  Re}\,\Pi_{_{L}}(0)
\end{equation}
In the HTL approximation, there is no screening for the transverse
static gluons, but this is not expected to hold generally. The
corresponding screening mass is the magnetic mass, and is denoted 
\begin{equation}
m_{\rm mag}^2\equiv {\rm
  Re}\,\Pi_{_{T}}(0)\; .
\end{equation}

In terms of those parameters, it is straightforward to write down the
expressions of $J_{_{T,L}}$ and $K_{_{T,L}}$ for photon production in a
description where we use gluon propagators that are more general than
the HTL propagators:
\begin{eqnarray}
&&J_{_{T}}=\pi\left[ F\Big({{4M_{\rm eff}^2}\over{m_{\rm mag}^2}}\Big) - 
F\Big({{4M_{\rm eff}^2}\over{m_P^2}}\Big)
\right]\; ,\nonumber\\
&&J_{_{L}}=\pi\left[ F\Big({{4M_{\rm eff}^2}\over{m_D^2}}\Big) - 
F\Big({{4M_{\rm eff}^2}\over{m_P^2}}\Big)
\right]\; ,\nonumber\\
&&K_{_{T}}=\pi\left[
-{1\over 8}\ln\Big({{m_{\rm mag}^2}\over{m_P^2}}\Big)
+{{M_{\rm eff}^2}\over{m_P^2}}F\Big({{4M_{\rm
eff}^2}\over{m_P^2}}\Big)
-{{M_{\rm eff}^2}\over{m_{\rm mag}^2}}F\Big({{4M_{\rm
eff}^2}\over{m_{\rm mag}^2}}\Big)
\right]\; ,\nonumber\\
&&K_{_{L}}=\pi\left[
-{1\over 8}\ln\Big({{m_D^2}\over{m_P^2}}\Big)
+{{M_{\rm eff}^2}\over{m_P^2}}F\Big({{4M_{\rm
eff}^2}\over{m_P^2}}\Big)
-{{M_{\rm eff}^2}\over{m_D^2}}F\Big({{4M_{\rm
eff}^2}\over{m_D^2}}\Big)
\right]\; .
\end{eqnarray}
It is easy to check that these relations fall back to
Eqs.~(\ref{eq:KL-final}) if we set $m_{\rm mag}=0$, $m_P=m_{\rm g}$
and $m_D=\sqrt{3} m_{\rm g}$, which are the relations between masses
in the HTL framework.

There is another general property of the processes of figure
\ref{fig:processes} which is worth mentioning here. Their rate in fact
depends only on the combinations $J_{_{T}}-J_{_{L}}$ and
$K_{_{T}}-K_{_{L}}$ (see Eq.~(\ref{eq:rate-final})) after one has summed
the contributions of transverse and longitudinal gluons. Using the above
formulas, we obtain:
\begin{eqnarray}
&&J_{_{T}}-J_{_{L}}=\pi\left[
 F\Big({{4M_{\rm eff}^2}\over{m_{\rm mag}^2}}\Big) -
 F\Big({{4M_{\rm eff}^2}\over{m_D^2}}\Big)\right]\; ,\nonumber\\
&&K_{_{T}}-K_{_{L}}=\pi\left[
{1\over 8}\ln\Big({{m_D^2}\over{m_{\rm mag}^2}}\Big)
+{{M_{\rm eff}^2}\over{m_D^2}}F\Big({{4M_{\rm
eff}^2}\over{m_D^2}}\Big)
-{{M_{\rm eff}^2}\over{m_{\rm mag}^2}}F\Big({{4M_{\rm
eff}^2}\over{m_{\rm mag}^2}}\Big)
\right]\; .
\end{eqnarray}
In other words, all the dependence on the plasmon mass drops out for the
processes of figure \ref{fig:processes}. This property is in fact
reasonable since we are looking at processes that involve only
space-like gluons, and it would have been surprising if the result had
depended on the plasmon mass, a property of time-like gluons. The
practical consequence of this for a phenomenological approach to photon
production based on some quasi-particle picture is that we do not need
to know the full gluon propagator, but only the two screening masses and
the quark thermal mass\footnote{This is not true for the
  $2\to 2$ processes calculated in \cite{AltheR1}. Indeed, since these
  processes involve time-like gluons, they can depend on the plasmon
  mass.}.  Note also that these quantities remain finite
even if the magnetic mass is very small or vanishing.

In particular, it is now known from lattice calculations
 \cite{PeshiKP1,PetreKLSW1} that the masses of quasi-particles increase
 when the temperature approaches the critical temperature from above,
 while at the same time the screening masses decrease.  The formulas of
 this section are important to deal with such a situation, since they do
 not assume any particular relationship between the screening masses and
 the quasi-particle masses.  For instance, for a temperature just above
 $T_c$, we can make use of Eqs.~(\ref{eq:Mggm}), and predict a
 suppression of the production rate of photons and low-mass dileptons
 simply due to the fact that screening masses are much smaller than the
 quasi-particle masses (in addition to the standard suppression due to
 the fact that the temperature is smaller).

\section{Resummation of ladder diagrams and LPM effect}
\label{sec:resum}
\subsection{Integral equation in momentum space}
The authors of \cite{ArnolMY1,ArnolMY2} performed the resummation of all
the ladder diagrams, as well as all the self-energy corrections that are
required to preserve gauge invariance, in order to account for the LPM
effect in the production of real photons. Indeed, such photons may have
a formation time larger than the mean free path of the quarks in the
medium \cite{AurenGZ2}, so that multiple quark scatterings
contribute coherently to the formation of the photon.

In the formulation of \cite{ArnolMY1,ArnolMY2}, the imaginary part of
the retarded photon polarization tensor is given by\footnote{Note that
  this equation includes only the two transverse modes of the photon,
  and is therefore incomplete for the production of virtual photons.
  It is however not difficult to include the longitudinal mode as well
  \cite{WorkIP1}.}
\begin{eqnarray}
&&\!\!\!\!{\rm Im}\,\Pi_{_{R}}{}_\mu^\mu(Q)\approx {{e^2
N_c}\over{8\pi}}
\int_{-\infty}^{+\infty}dp_0\,
\;[n_{_{F}}(r_0)-n_{_{F}}(p_0)]\;
{{p_0^2+r_0^2}\over{p_0^2r_0^2}}
\nonumber\\
&&\qquad\qquad\qquad\times{\rm Re}\,\int {{d^2{\imb p}_\perp}\over{(2\pi)^2}}\;
{\imb
p}_\perp\cdot{\imb f}({\imb p}_\perp)\; ,\nonumber\\
&&
\label{eq:LPM}
\end{eqnarray}
where $r_0\equiv p_0+q_0$ and ${\imb f}({\imb p}_\perp)$ is a
dimensionless transverse vector that represents the resummed coupling of
a quark to the transverse modes of the photon, satisfying the following
integral equation
\begin{equation}
2{\imb p}_\perp=i\delta E {\imb f}({\imb p}_\perp)+g^2 C_{_{F}}T\int
{{d^2{\imb l}_\perp}\over{(2\pi)^2}} {\cal C}({\imb l}_\perp)
[{\imb f}({\imb p}_\perp)-{\imb f}({\imb p}_\perp+{\imb l}_\perp)]\; ,
\label{eq:integ-f}
\end{equation}
where $\delta E\equiv q_0({\imb p}_\perp^2+M_{\rm eff}^2)/(2p_0r_0)$
and in which ${\cal C}({\imb l}_\perp)$ is the collision integral defined by
\begin{eqnarray}
&&{\cal C}({\imb l}_\perp)\equiv
\int
{{dl_0dl_z}\over{(2\pi)^2}} 2\pi\delta(l_0-l_z)\,{1\over{l_0}}\nonumber\\
&&\qquad\quad\times
\sum_{\alpha=L,T}  {{2{\rm Im}\,\Pi_{\alpha}(L)}\over{(L^2-{\rm
Re}\,\Pi_{\alpha}(L))^2+({\rm Im}\,\Pi_{\alpha}(L))^2}}\;
P_{\alpha}^{\mu\nu}(L) \widehat{Q}_\mu \widehat{Q}_\nu\; ,
\label{eq:coll-term}
\end{eqnarray}
with $\widehat{Q}_\mu\equiv (1,{\imb q}/q)$ and
$P_{_{T,L}}^{\mu\nu}(L)$ the transverse or longitudinal projector for
a gluon of momentum $L$.  Note that $\delta E^{-1}$ is nothing but the
typical formation time of the photon \cite{AurenGZ2}.
Note also that in an iterative solution of these integral equations,
the first term that contributes to the imaginary part of the photon
polarization tensor is of order $g^2$ since ${\imb f}({\imb p}_\perp)$
and $g({\imb p}_\perp)$ are purely imaginary at the order $g^0$. This
reflects the fact that the direct production of a photon by the
processes $q\bar{q}\to\gamma$ or $q\to q\gamma$ is kinematically
forbidden. This remark ceases to be valid if $\delta E$ can vanish, in
which case an $i\epsilon$ prescription must be used, so that the
zeroth order solution can have a real part. This happens if $M_{\rm
  eff}^2$ can become negative, i.e. if $Q^2>4M_\infty^2$.

Using the fact that we have\footnote{One can use
\begin{equation}
P_{_{T}}^{\mu\nu}(L)+P_{_{L}}^{\mu\nu}(L)=g^{\mu\nu}-{{L^\mu L^\nu}\over{L^2}}\; ,
\end{equation}
and 
\begin{equation}
\widehat{Q}_\mu \widehat{Q}_\nu\left(g^{\mu\nu}-{{L^\mu L^\nu}\over{L^2}}\right)=0 \quad{\rm if}\quad {l_0=l_z}
\end{equation}
in order to obtain the second contraction.}
\begin{equation}
P_{_{T}}^{\mu\nu}(L) \widehat{Q}_\mu \widehat{Q}_\nu=
{{l_z^2}\over{l^2}}-1 = 
-P_{_{L}}^{\mu\nu}(L) \widehat{Q}_\mu \widehat{Q}_\nu\; ,
\end{equation}
and introducing the variable $x\equiv l_0/l$, we can
rewrite\footnote{This change of coordinates has the following Jacobian:
\begin{equation}
\left(1-{{l_0^2}\over{l_0^2+l_\perp^2}}\right) {{dl_0}\over{l_0}}
d(l_\perp^2)= {{dx}\over x}d(l_\perp^2)\; .
\end{equation}} the collision integral in Eq.~({\ref{eq:coll-term}}) as
\begin{eqnarray}
&&\!\!\!\!{\cal C}({\imb l}_\perp)={2\over\pi} \int_0^1{{dx}\over {x}}
\;\left[
{{{\rm Im}\,\Pi_{_{L}}(x)}\over{(l_\perp^2+{\rm
Re}\,\Pi_{_{L}}(x))^2+({\rm Im}\,\Pi_{_{L}}(x))^2}}\right.\nonumber\\
&&\qquad\qquad\qquad\qquad\left.
-
{{{\rm Im}\,\Pi_{_{T}}(x)}\over{(l_\perp^2+{\rm
Re}\,\Pi_{_{T}}(x))^2+({\rm Im}\,\Pi_{_{T}}(x))^2}}
\right]\; .
\end{eqnarray}
At this point, the sum rule derived in section \ref{sec:sum-rule} gives
directly the result of the integral over the variable $x$, so that the
collision integral 
can be rewritten as:
\begin{equation}
{\cal C}({\imb l}_\perp)={1\over{l_\perp^2}}-{1\over{l_\perp^2+3m_{\rm g}^2}}\; .
\end{equation}
We observe again that summing over the contributions of transverse and
longitudinal gluon exchanges cancels the terms involving the plasmon
mass.

\subsection{Solution in the Bethe-Heitler regime}
As a check, one can solve this integral equation iteratively up to the
order $g^2$. Indeed, the naive term of order $g^0$ is purely imaginary,
and drops out of the imaginary part of the photon polarization tensor
(see Eq.~(\ref{eq:LPM})). For the function ${\imb f}({\imb
p}_\perp)$, this expansion gives:
\begin{eqnarray}
&&\!\!\!\!\!\!{\rm Re}\;\int {{d^2{\imb p}_\perp}\over{(2\pi)^2}}\; {\imb p}_\perp\cdot {\imb f}({\imb p}_\perp)
\empile{=}\over{{\cal O}(g^2)}8g^2 C_{_{F}}T {{(p_0r_0)^2}\over{q_0^2}}
\int {{d^2{\imb p}_\perp}\over{(2\pi)^2}}\;
{{{\imb p}_\perp}\over{{\imb p}_\perp^2+M_{\rm eff}^2}}\nonumber\\
&&\qquad\qquad\cdot
\int {{d^2{\imb l}_\perp}\over{(2\pi)^2}} 
{\cal C}({\imb l}_\perp)
\left[{{{\imb p}_\perp}\over{{\imb p}_\perp^2+M_{\rm eff}^2}}
-{{{\imb p}_\perp+{\imb l}_\perp}\over{({\imb p}_\perp+{\imb l}_\perp)^2+M_{\rm eff}^2}}
\right]\; .\nonumber\\
&&
\end{eqnarray}
At this stage, performing the angular integrations is elementary.
Making use of the following identity:
\begin{equation}
\int\limits_0^{+\infty}d(p_\perp^2)\left(
{1\over{p_\perp^2+M_{\rm eff}^2}}
-{1\over{\sqrt{(p_\perp^2+l_\perp^2+M_{\rm eff}^2)^2-4p_\perp^2 l_\perp^2}}}
\right)=0\; ,
\end{equation}
this can be rewritten as:
\begin{eqnarray}
{\rm Re}\;\int {{d^2{\imb p}_\perp}\over{(2\pi)^2}}\; {\imb p}_\perp\cdot {\imb f}({\imb p}_\perp)
\empile{=}\over{{\cal O}(g^2)}
-{{2g^2C_{_{F}}T}\over{\pi^3}}{{(p_0r_0)^2}\over{q_0^2}}[J_{_{T}}-J_{_{L}}+2K_{_{T}}-2K_{_{L}}]\; .
\label{eq:f-BH}
\end{eqnarray}
Finally, plugging Eqs.~(\ref{eq:f-BH}) into
Eq.~(\ref{eq:LPM}) gives
\begin{eqnarray}
&&\!\!\!\!{\rm Im}\,\Pi_{_{R}}{}_\mu^\mu(Q)
\empile{=}\over{{\cal O}(g^2)}
-{{e^2 g^2 N_c C_{_{F}}}\over{2\pi^4}}{T\over{q_0^2}}
\int\limits_{-\infty}^{+\infty}dp_0[n_{_{F}}(r_0)-n_{_{F}}(p_0)]\nonumber\\
&&\qquad\times\,
(p_0^2+r_0^2)[J_{_{T}}-J_{_{L}}+2K_{_{T}}-2K_{_{L}}]
\; ,
\label{eq:rate-g2}
\end{eqnarray}
which is equivalent to Eq.~(\ref{eq:rate-final}) for real photons
($Q^2=0$, $M_{\rm eff}^2=M_\infty^2$). This proves the agreement between
the perturbative approach followed in \cite{AurenGZ3} and the resummation of
the LPM corrections, if one formally keeps only the ${\cal O}(g^2)$
terms. The fact that some terms proportional to $Q^2$ in
Eq.~(\ref{eq:rate-final}) are not recovered in this limit is due to the
fact that the LPM resummation of \cite{ArnolMY1,ArnolMY2} is limited to
the transverse modes of the produced photon, while massive photons also
have a physical longitudinal mode.

\subsection{Reformulation as a differential equation}
\label{sec:diff}
Since the collision integral appearing under the integral over ${\imb
  l}_\perp$ is now known in closed form, it is possible to transform
this integral equation into an ordinary differential equation by going
to impact parameter space. We can first define
\begin{equation}
{\imb f}({\imb p}_\perp)\equiv\int d^2{\imb b} \,
 e^{-i{\imb p}_\perp\cdot{\imb b}}
{\imb g}({\imb b})\; .
\end{equation}
Note that the order zero (in $g^2$) solution of the integral equation
is
\begin{equation}
{\imb g}_0({\imb b})=-{2\over \pi} {{p_0r_0}\over{q_0}} \nabla_b K_0(M_{\rm eff} b)={2\over \pi} {{p_0r_0}\over{q_0}} M_{\rm eff} \widehat{\imb b} K_1(M_{\rm eff} b)\; ,
\end{equation}
where the $K_i$'s are modified Bessel functions of the second kind.
For the higher order terms ${\imb g}_1({\imb b})$, one obtains the
following equation for ${\imb g}_1$
\begin{equation}
i{{q_0}\over{2p_0r_0}}(M_{\rm eff}^2-\Delta_\perp){\imb g}_1({\imb
b})+g^2C_{_{F}}TD(m_{\rm g}b)({\imb g}_0({\imb b})+{\imb g}_1({\imb b}))=0\; ,
\end{equation}
with
\begin{equation}
D(m_{\rm g}b)\equiv {1\over{2\pi}}\left[
\gamma+\ln\left({{\sqrt{3}m_{\rm g} b}\over{2}}\right)
+K_0(\sqrt{3}m_{\rm g}b)
\right]\; .
\end{equation}
In addition, we have also
\begin{equation}
{\rm Re}\; {\imb p}_\perp \cdot \int {{d^2{\imb p}_\perp}\over{(2\pi)^2}}
{\imb f}({\imb p}_\perp)=\lim_{b\to 0^+}{\rm Im}\;
\nabla_\perp\cdot{\imb g}_1({\imb b})\; .
\end{equation}

This differential equation can be further simplified by defining the
following dimensionless quantities:
\begin{eqnarray}
&&t\equiv M_{\rm eff}^2 b^2\; ,\nonumber\\
&&u(t)\equiv {{\pi q_0}\over{2p_0 r_0}} {\imb b}\cdot {\imb g}({\imb b})\; .
\end{eqnarray}
This transformation leads to
\begin{equation}
4t u_1^{\prime\prime}(t)-u_1(t)+ig^2C_{_{F}} {{p_0
r_0}\over{q_0}}{T\over{M_{\rm eff}^2}} D\left({{m_{\rm g}}\over{M_{\rm
eff}}}\sqrt{t}\right)(u_0(t)+u_1(t))=0\; ,
\label{eq:diff}
\end{equation}
where the prime denotes the differentiation with respect to $t$, and where
\begin{equation}
u_0(t)\equiv \sqrt{t}K_1(\sqrt{t})\; .
\end{equation}
The differential equation Eq.~(\ref{eq:diff}) depends on two
dimensionless quantities. One is the ratio $m_{\rm g}/M_{\rm eff}$ of
two masses, while the prefactor $g^2C_{_{F}}T p_0 r_0/q_0M_{\rm eff}^2$
can be interpreted (up to logarithms) as the ratio of the photon
formation time to the quark mean free path. It is therefore the average
number of scatterings that can contribute coherently to the production
of a photon. The relevant quantity for the photon production rate is
then given by
\begin{equation}
{\rm Re}\;{\imb p}_\perp \cdot \int {{d^2{\imb p}_\perp}\over{(2\pi)^2}}
{\imb f}({\imb p}_\perp)= {{4p_0r_0}\over{\pi q_0}}M_{\rm eff}^2 \;{\rm Im}\;
u_1^\prime(0)\; .
\end{equation}

This differential equation must be supplemented by boundary conditions
in order to define uniquely the solution. First, we have
\begin{equation}
u(0)=0\; ,
\end{equation}
which can be seen from its definition. Subtracting $u_0(0)=1$, this implies
\begin{equation}
u_1(0)=-1 \; .
\end{equation}
Unfortunately, we do not know the value of $u_1^\prime(0)$ but instead we know the value of $u(\infty)$. Indeed,
assuming that ${\imb f}({\imb p}_\perp)$ is regular enough, its
Fourier transform is exponentially suppressed at large $b$. Therefore,
$u(\infty)=0$. Since we already have $u_0(\infty)=0$, this implies
\begin{equation}
u_1(\infty)=0\; .
\end{equation}
Therefore, the problem can be summarized as follows: we are looking for
the (presumably complex) value of $u_1^\prime(0)$ that takes us from
$u_1(0)=-1$ to $u_1(\infty)=0$. The imaginary part of this derivative
term is then the coefficient that enters in the photon polarization
tensor.  Note also that since the term $u_1^{\prime\prime}$ and $u_1$
come with opposite signs in the combination $4tu_1^{\prime\prime} -u_1$
in Eq.~(\ref{eq:diff}), generic solutions\footnote{The generic solution
of $4tu_1^{\prime\prime}(t)-u_1(t)=0$ is $u_1(t)=\sqrt{t}(c_1
I_1(\sqrt{t})+c_2 K_1(\sqrt{t}))$, where $I_1$ is a modified Bessel
function of the first kind.} are unstable and diverge as $t\to +\infty$
(they are in fact linear combinations of a function that diverges
exponentially and of a function that goes exponentially to zero). It is
only for a very specific value of $u_1^\prime(0)$ that one can make the
coefficient of the diverging term zero, and have a solution that
satisfies $u_1(\infty)=0$. This heuristic argument thereby justifies the
uniqueness of the number $u_1^\prime(0)$ solving the problem.

\subsection{Numerical solution}
\subsubsection{General principle}
This reformulation of Eqs.~(\ref{eq:LPM}) and (\ref{eq:integ-f}) leads
to a rather straightforward algorithm for a numerical solution.
Differential problems with two-point boundary conditions are usually
solved by iterative ``shooting'' methods \cite{NR}, where one tries to make a
guess for the missing initial condition (here, $u_1^\prime(0)$) and
then corrects this value based on the discrepancy between the
resulting end-point and the expected one.

Things are in fact much simpler for an affine equation, since
only two trials, plus the knowledge of a solution of the complete
equation, are enough to find the value of $u_1^\prime(0)$. Indeed, for
a differential equation like (\ref{eq:diff}), solutions have
generically the following form:
\begin{equation}
u_1(t)=w(t)+\alpha_1 w_1(t) + \alpha_2 w_2(t)\; ,
\end{equation}
where $w(t)$ is a particular solution of the full equation, and
$w_{1,2}(t)$ are two independent solutions of the corresponding
homogeneous equation (i.e. (\ref{eq:diff}) in which one would set
$u_0$ to zero). Generically, it is convenient to chose these functions
such that:
\begin{eqnarray}
&&w(0)=0\; ,\qquad w^\prime(0)=0\; ,\nonumber\\
&&w_1(0)=1\; ,\qquad w_1^\prime(0)=0\; ,\nonumber\\
&&w_2(0)=0\; ,\qquad w_2^\prime(0)=1\; .
\label{eq:naive-init-cond}
\end{eqnarray}
If these solutions are known (at least numerically), the
condition $u_1(0)=-1$ implies
\begin{equation}
\alpha_1=-1\; .
\end{equation}
Then, using $u_1(+\infty)=0$, the second coefficient $\alpha_2$ is
given by
\begin{equation}
\alpha_2=\lim_{t\to+\infty}{{w_1(t)-w(t)}\over{w_2(t)}}\; ;
\end{equation}
and thanks to our choice of initial conditions for $w,w_1$ and $w_2$,
the value of $u_1^\prime(0)$ that solves the problem is simply given
by
\begin{equation}
u_1^\prime(0)=\alpha_2\; .
\end{equation}

\subsubsection{Realistic algorithm}
This algorithm is however not directly applicable in the case of
Eq.~(\ref{eq:diff}), because the point $t=0$ is a singular point of
the equation, and cannot be used to set initial conditions. Therefore,
we have to chose another point in order to set the initial conditions.
Let us call this point $t_0>0$, and assume
\begin{eqnarray}
&&w(t_0)=0\; ,\qquad w^\prime(t_0)=0\; ,\nonumber\\
&&w_1(t_0)=1\; ,\qquad w_1^\prime(t_0)=0\; ,\nonumber\\
&&w_2(t_0)=0\; ,\qquad w_2^\prime(t_0)=1\; ,
\end{eqnarray}
instead of Eqs.~(\ref{eq:naive-init-cond}). From these initial
conditions, one must evolve numerically the functions $w,w_1$ and
$w_2$ both forward and backward. Having in mind what has been said at
the end of section \ref{sec:diff}, the three functions are going to
diverge when $t\to+\infty$ (unless one has been very unlucky when
choosing the initial conditions). The condition $u_1(+\infty)=0$ then
implies
\begin{equation}
0=1+\alpha_1 r_1 +\alpha_2 r_2\; ,
\end{equation}
where $r_i\equiv \lim_{t\to+\infty} w_i(t)/w(t)$. This gives a first
linear relation between the two coefficients $\alpha_1$ and
$\alpha_2$. Then, from the condition $u_1(0)=-1$, one can obtain
\begin{equation}
\alpha_1=\lim_{t\to 0^+}-{{w_2(t)-r_2(1+w(t))}\over{r_1 w_2(t)-r_2 w_1(t)}}\; .
\end{equation}
At this point, the two coefficients $\alpha_{1,2}$ are known, and it
is easy to find
\begin{equation}
u_1^\prime(0)=\lim_{t\to 0^+} w^\prime(t)+\alpha_1 w_1^\prime(t) + \alpha_2 w_2^\prime(t)\; .
\end{equation}
This last step does not require any additional calculation, since the
derivatives of $w,w_{1,2}$ are known numerically at this point. In
summary, this algorithm reduces the problem of solving the integral
equation (\ref{eq:integ-f}) and then calculating the integral over
${\imb p}_\perp$ in Eq.~(\ref{eq:LPM}) to the numerical solution of a
differential equation with three different initial conditions. A
numerical analysis of Eq.~({\ref{eq:LPM}}) based along these lines
will be presented elsewhere \cite{WorkIP1}.

\section{Conclusions}
In this paper, we have derived a simple sum rule that enables to perform
analytically some of the integrals involved in the thermal 2-loop photon
production rate. Several applications of this sum rule have been
presented. This sum rule also plays a role in the calculation of the
collision integral that appears in the resummation of the ladder diagrams
involved in the calculation of the LPM effect. 

\section*{Acknowledgments}
F.G. would like to thank E. Fraga and A. Peshier for many useful
discussions on related issues. H.Z. thanks the LAPTH for hospitality
during the summer of 2001, where part of this work has been performed.
P. A. thanks C. Gale for the hospitality extended to him at Mc Gill
University where part of this work was done.

\appendix

\section{Asymptotic behavior of $F(x)$}
\label{app:F-asympt}
\subsection{Large $x$ behavior of $F(x)$}
At large values of the argument $x$, the asymptotic value of the
function $F(x)$ introduced in Eq.~(\ref{eq:F-def}) is very easy to
obtain\footnote{By subtracting more of the Taylor expansion of ${\rm
tanh}^{-1}(u)$, one can go one order further in this asymptotic expansion,
and obtain
\begin{equation}
F(x)\empile{\approx}\over{x\to\infty}{{\ln(x)}\over{2x}}+{{1-\ln(2)}\over
x}
+{{\ln(x)}\over{3x^2}}+{{5-6\ln(2)}\over{9x^2}}+{\cal O}\Big({{\ln(x)}\over{x^3}}\Big)\; .
\end{equation}}:
\begin{eqnarray}
F(x)&=&\epsilon \int_0^1 du {{{\rm
tanh}^{-1}(u)}\over{u^2+\epsilon}}\qquad{\rm with}\qquad \epsilon\equiv
{1\over{x-1}}\ll 1\nonumber\\
&=&\epsilon\int_0^1du {{{\rm tanh}^{-1}(u)-u}\over{u^2+\epsilon}}
+\epsilon\int_0^1du
{u\over{u^2+\epsilon}}\nonumber\\
&=&\epsilon\left[1-\ln(2)+{\cal O}(\epsilon\ln(1/\epsilon))\right]
+\epsilon\ln\Big({{1+\epsilon}\over\epsilon}\Big)\; ,
\end{eqnarray}
i.e.
\begin{eqnarray}
F(x)\empile{=}\over{x\to+\infty} {{\ln(x)}\over{2x}}+{{1-\ln(2)}\over x}
+{\cal O}\Big({{\ln(x)}\over{x^2}}\Big)\; .
\end{eqnarray}
Let us add that if the variable $x$ goes to $-\infty$, one can obtain
the correct asymptotic behavior by replacing $x$ by $-x$ in the previous
expression (in particular $\ln(x)$ by $\ln|x|-i\pi$) while dropping the
imaginary part, so that the previous asymptotic formula simply becomes:
\begin{equation}
F(x)\empile{=}\over{x\to-\infty} {{\ln|x|}\over{2x}}+{{1-\ln(2)}\over x}
+{\cal O}\Big({{\ln|x|}\over{x^2}}\Big)\; .
\end{equation}

\subsection{Small $x$ behavior of $F(x)$}
At small values of $x$, determining the expansion of $F(x)$ requires a
little more work. Denoting $y\equiv 1/\sqrt{1-x}\approx 1+x/2$, we have
\begin{eqnarray}
F(x)&=& {{y^2}\over 2}\int_0^1 du
\ln\Big({{1+u}\over{1-u}}\Big){1\over{y^2-u^2}}\nonumber\\
&=& {y\over 4} \int_0^1 du
\ln\Big({{1+u}\over{1-u}}\Big) \left[ {1\over {y+u}}+{1\over{y-u}}\right]\; .
\end{eqnarray}
Let us notice first that the term in $1/(y+u)$ is finite in the limit
$y\to 1$. Therefore, since we do not want to go beyond the constant
terms, we can simply replace $y$ by $1$ in this term and get
\begin{equation}
{1\over 4}\int\limits_0^1 du
\ln\Big({{1+u}\over{1-u}}\Big){1\over {1+u}}=-{1\over 4}\int\limits_0^1 dv {{\ln(v)}\over{1+v}}={{\pi^2}\over{48}}\; .
\end{equation}
The other term can be separated in two parts, the first one being
\begin{eqnarray}
{y\over 4}\int\limits_0^1du {{\ln(1+u)}\over{y-u}}
&=&{y\over4}\int\limits_0^1 {{\ln(1+u)-\ln(2)}\over{y-u}}+{y\over 4}\int\limits_0^1 {{\ln(2)}\over{y-u}}\nonumber\\
&\approx& {1\over 4}\int\limits_0^1 {{dv}\over v}\ln(1-v/2)+{{\ln(2)}\over 4}\ln\Big({2\over x}\Big)\nonumber\\
&\approx&-\int\limits_0^{1/2} {{dv}\over 4}{{\ln(1-v)}\over{1-v}}
+\int\limits_0^{1/2} {{dv}\over 4}{{\ln(1-v)}\over{v(1-v)}}+{{\ln(2)}\over 4}\ln\Big({2\over x}\Big)\nonumber\\
&\approx& {{\ln^2(2)}\over 8}-{{\pi^2}\over{48}}+{{\ln(2)}\over 4}\ln\Big({2\over x}\Big)\; ,
\end{eqnarray}
up to terms that vanish when $x\to 0^+$.
Seemingly, the second piece is:
\begin{eqnarray}
-{y\over 4}\int\limits_0^1du {{\ln(1-u)}\over{y-u}}&=&
-{y\over 4}\int\limits_0^1 du {{\ln(y-u)+\ln(1-{{y-1}\over{y-u}})}\over{y-u}}
\nonumber\\
&\approx&{1\over 8}\ln^2\Big({2\over x}\Big)+{1\over 4}\sum_{n=1}^{+\infty} {{(y-1)^p}\over{p}}
\int\limits_0^1 {{du}\over{(y-u)^{p+1}}}\nonumber\\
&\approx&{1\over 8}\ln^2\Big({2\over x}\Big)+{{\pi^2}\over{24}}\; .
\end{eqnarray}
Collecting all the bits and pieces, we finally obtain:
\begin{equation}
F(x)\empile{=}\over{x\to 0^+} {1\over 8} \ln^2\Big({4\over x}\Big)+{{\pi^2}\over{24}}+{\cal O}\Big(x\ln^2(1/x)\Big)\; .
\end{equation}
When $x$ approaches $0$ by negative values, it is sufficient to replace
$x$ by $-x$ in the previous formula (i.e. $\ln(x)$ by $\ln|x|-i\pi$) and
retain only the real part. Since the logarithm is squared, the $-i\pi$
modifies the constant term as follows:
\begin{equation}
F(x)\empile{=}\over{x\to 0^-} {1\over 8} \ln^2\Big|{4\over
x}\Big|-{{\pi^2}\over{12}}+{\cal O}\Big(x\ln^2(1/x)\Big)\; .
\label{eq:F-small-neg-x}
\end{equation}

\section{For real photons: $J_T=-J_L$ if $M_\infty=m_{\rm g}$}
\label{app:B}
In the production of real photons, one can notice numerically that the
coefficients $J_{_{T}}$ and $J_{_{L}}$ are equal if the gluon plasmon
mass $m_{\rm g}$ is equal to the quark asymptotic mass $M_\infty$
\cite{Kobes4} (for real photons $M_{\rm eff}=M_\infty$). We present
here an analytic proof of this statement. When $M_\infty=m_{\rm g}$ and
$Q^2=0$, we have:
\begin{eqnarray}
J_{_{L}}+J_{_{T}}&=&\pi\left[F(4/3)-2F(4)\right]\nonumber\\
&=&\pi\left[
3\int\limits_0^1 du{{{\rm tanh}^{-1}(u)}\over{u^2+3}}
-2\int\limits_0^1 du{{{\rm tanh}^{-1}(u)}\over{3u^2+1}}
\right]\; .
\end{eqnarray}
Noticing that the function 
\begin{equation}
G(x)\equiv \int_0^1 du {{{\rm tanh}^{-1}(u)}\over{u^2+x^2}}
\end{equation}
obeys the differential equation
\begin{equation}
G(x)+xG^\prime(x)={1\over
2}{{\ln\Big({{1+x^2}\over{4x^2}}\Big)}\over{1+x^2}}\; ,
\end{equation}
and solving it, one can prove the following formula
\begin{eqnarray}
x\int_0^1 du {{{\rm tanh}^{-1}(u)}\over{u^2+x^2}}&=&\ln(2){\rm
tan}^{-1}\Big({1\over x}\Big) -{1\over 2}\int_0^{1/x}du
{{\ln(1+u^2)}\over{1+u^2}}\nonumber\\
&=&\ln(2){\rm
tan}^{-1}\Big({1\over x}\Big)+\int_0^{{\rm tan}^{-1}(1/x)} d\theta
\ln(\cos(\theta))\; .
\end{eqnarray}
Using this intermediate result, it is easy to rewrite
$J_{_{T}}+J_{_{L}}$ as
\begin{equation}
J_{_{T}}+J_{_{L}}={\pi\over{\sqrt{3}}}\left[
3\int_0^{\pi/6}d\theta \ln(\cos(\theta))
-2\int_0^{\pi/3}d\theta \ln(\cos(\theta))
-{{\pi\ln(2)}\over 6}
\right]\; .
\end{equation}
Making then use of the following relations
\begin{eqnarray}
&&\int_0^{\pi/2}d\theta \ln(\cos(\theta))=-{{\pi\ln(2)}\over 2}\;
,\nonumber\\
&&\int_0^{\pi/3}d\theta \ln(\cos(\theta))=-{{\pi\ln(2)}\over 2}
-\int_0^{\pi/6}d\theta \ln(\sin(\theta))\; ,\nonumber\\
&&\int_0^{\pi/6}d\theta \ln(\cos(\theta))=-{{\pi\ln(2)}\over 2}
-\int_0^{\pi/3}d\theta \ln(\sin(\theta))\; ,\nonumber\\
&&\int_0^{\pi/6}d\theta \ln(\sin(\theta))+\int_0^{\pi/6}d\theta
\ln(\cos(\theta))={1\over 2}\int_0^{\pi/3}d\theta \ln(\sin(\theta))
-{{\pi\ln(2)}\over 6}\; ,\nonumber\\
&&
\end{eqnarray}
it is straightforward to check that
\begin{equation}
J_{_{T}}+J_{_{L}}=0
\end{equation}
when $M_\infty=m_{\rm g}$. This is an interesting non trivial (and exact) identity
which cannot be obtained by any simple other method, mainly because there
is no useful asymptotic formula near the point $M_\infty=m_{\rm g}$. Even if
this identity is purely anecdotical, its derivation
illustrates the power of the sum rule obtained in this paper.

\bibliographystyle{unsrt}

\end{document}